\documentclass[useAMS,referee]{biom}
\usepackage{algorithm}
\usepackage{algpseudocode}
\usepackage{graphicx}
\usepackage{xcolor}
\usepackage{subcaption}
\let\bibhang\relax
\usepackage{natbib}
\usepackage{amsmath} 
\usepackage{url}

\def\bSig\mathbf{\Sigma}

\title{LoopPerm-CPD: A Robust Loop Permutation Framework for Automatic Multiple Change-Point Detection in Longitudinal Data}

\author{
Xuejun Sun$^{1}$, Oliver Li$^{1}$, Qianhui Zheng$^{2}$, Xiaojing Zheng$^{1,3,*}$\email{xiaojinz@email.unc.edu}, and Fei Zou$^{1,4,**}$\email{feizou@email.unc.edu} \\
$^{1}$Department of Biostatistics, \\
University of North Carolina at Chapel Hill, Chapel Hill, North Carolina, U.S.A. \\
$^{2}$Duke Global Health Institute, Duke University, Durham, North Carolina, U.S.A. \\
$^{3}$Department of Pediatrics, \\
University of North Carolina at Chapel Hill, Chapel Hill, North Carolina, U.S.A.\\
$^{4}$Department of Genetics, \\
University of North Carolina at Chapel Hill, Chapel Hill, North Carolina, U.S.A.
}

\begin{document}


\date{{\it Received October} 2007. {\it Revised February} 2008.  {\it
Accepted March} 2008.}



\pagerange{\pageref{firstpage}--\pageref{lastpage}} 
\volume{64}
\pubyear{2026}
\artmonth{June}



\label{firstpage}


\begin{abstract}
Human viral challenge studies --- in which participants are deliberately inoculated with influenza strains such as H1N1 or H3N2 \citep{liu2016individualized} and monitored through longitudinal transcriptomic profiling before and after inoculation --- are critical for characterizing dynamic biological immune responses to these viruses. A key analytical goal in such settings is detecting critical transition times, or change points, at which an underlying trajectory shifts direction or rate, indicating, for example, the onset of an immune response or recovery. Yet change-point detection in these longitudinal data is fundamentally difficult --- observations are often sparse and irregularly spaced, sample sizes are small, outliers are common, and the number of change points is unknown a priori.

To address these challenges, we propose LoopPerm-CPD: a robust change-point detection approach with a built-in loop permutation procedure for automatic multiple change-point detection. The method evaluates candidate slope change points and assesses their significance via a novel within-subject circular permutation combined with a binary segmentation procedure, jointly estimating both the number and locations of change points. The R package \texttt{LoopPerm-CPD}, which implements this loop permutation framework, flexibly accommodates generalized least squares, quantile regression, and quantile rank-score statistics for various types of outcome variables.

We validate the proposed approach through simulations, demonstrating Type I error control and power superior to competing methods. Applied to real data, the framework recovers interpretable transition points in multiple human respiratory viral inoculation studies. Together, these results establish the method and its companion software as a robust and user-friendly tool for change-point detection in complex human longitudinal cohort data.

\end{abstract}



\maketitle


%

\section{INTRODUCTION}
\label{s:intro}
Longitudinal studies have been conducted across a wide range of human complex diseases, such as respiratory viral infections \citep{sun2025hrvilage3k3m}, to characterize the dynamic biological processes underlying these diseases. Driven by community efforts, longitudinal transcriptomic data from these studies are becoming increasingly available in public repositories such as GEO \citep{barrett2012geo}, opening new opportunities for systematic analysis of how gene-expression trajectories evolve over time. These analyses span applications from host immune dynamics following viral challenge to developmental processes such as lineage commitment, tissue differentiation, and stage-specific shifts in gene regulation \citep{cheemarla2021dynamic_innate,ratnasiri2024systems_immunology}. Related trajectory-based analyses have been used to study transitions between developmental states in biological systems using single-cell gene-expression data \citep{schiebinger2019optimal}. However, such studies are often small, with limited sample sizes, sparse and irregularly spaced time points, and outliers, posing substantial challenges for statistical analysis of gene-expression trajectories.

A key analytical goal in these settings is the identification of change points, or critical transition times at which an underlying trajectory shifts direction or rate, marking biologically meaningful events such as the onset of an immune response or a switch in developmental state. Change-point detection provides a principled statistical framework for identifying such transitions. Existing methods span a broad range of strategies, including likelihood-ratio tests, cumulative sum (CUSUM) procedures, and segmentation-based approaches \citep{page1954continuous,hinkley1971inference,chen1997testing,truong2020selective}. A major challenge in applying these classical approaches to longitudinal data is that they are mainly developed for independent observations 
and may not be directly appropriate for longitudinal data with repeated measurements \citep{vostrikova1981detecting,bai1997estimating,fryzlewicz2014wild}.

Several methods have been developed specifically for longitudinal change-point analysis. For example, \citet{xing2012semiparametric} proposed a semiparametric change-point regression model for longitudinal observations, \citet{lai2014identifying} developed a linear mixed-effects framework for identifying multiple change points, and \citet{fiecas2024generalised} proposed generalized mixed-effects models for change-point analysis of biomedical time-series data. However, many of these approaches rely on distributional assumptions and perform model-based likelihood inference with asymptotic approximations, which may be less reliable for studies with limited sample sizes.

Another important challenge is that the number of change points is usually unknown a priori. Some methods require the number of change points to be specified or selected in advance, such as \texttt{mcp} \citep{lindelov2020mcp}, \texttt{strucchange::breakpoints} \citep{zeileis2002strucchange}, and linear mixed-effects multiple change-point models \citep{lai2014identifying}. whereas others estimate it using penalized criteria or recursive search procedures. Dynamic programming and binary segmentation have been widely used to estimate both the number and locations of multiple change points \citep{bai1997estimating,fryzlewicz2014wild,killick2012optimal}.

Formal hypothesis testing for the null hypothesis of no change point has also been studied extensively, though largely outside the longitudinal setting. For example, \citet{andrews1993tests} developed Wald, Lagrange multiplier, and likelihood-ratio-type tests for parameter instability with an unknown change point, where the change-point parameter is present only under the alternative and leads to nonstandard asymptotic distributions. This nonregularity is related to the broader problem of nuisance parameters that are not identified under the null hypothesis \citep{hansen1996inference}. Ratio-based CUSUM tests have also been proposed to avoid direct estimation of nuisance scale parameters and improve applicability under dependent errors \citep{horvath2008ratio}. Nevertheless, most of these testing frameworks do not directly address repeated-measures longitudinal data with within-subject dependence.

Permutation methods provide an alternative route to statistical inference without relying entirely on asymptotic approximations. Existing permutation-based change-point procedures have been used to approximate null distributions and critical values for structural change tests in settings where the data are treated as independent observations or as a single time series, rather than repeated measurements nested within subjects \citep{antoch2001permutation,huskova2001permutation,zeileis2013permutation,cabrieto2018permutation}. A key limitation is that many of these procedures permute raw observations directly, which is likely to violate the exchangeability assumption when applied to irregularly spaced longitudinal data, leading to an invalid null distribution, and requiring an modified permutation that largely maintain the exchangeability of the data. 

These challenges motivate a robust finite-sample testing framework that can automatically estimate the number of change points, account for within-subject dependence, and provide formal statistical confidence for detected changes without relying entirely on large-sample approximations for sparse and irregularly spaced longitudinal data with potentially contaminated outliers. We propose the Loop Permutation framework, which evaluates candidate change points using slope-change models. Statistical significance is assessed by fitting a null model without change points and circularly permuting residual trajectories within each subject, thereby preserving local within-subject dependence while generating permutation samples under the no-change-point null. The framework is flexible and supports multiple working models and test statistics, including generalized least squares (GLS), quantile regression, and quantile rank-score tests. Embedded within binary segmentation, Loop Permutation automatically estimates both the number and locations of change points. The remainder of this paper is organized as follows: Section~2 introduces the proposed methodology, Section~3 presents simulation studies, Section~4 illustrates real-data applications, and Section~5 concludes with a discussion.

\section{METHODOLOGY}
\label{s:model}
\subsection{Setup and Notation}
\label{sec:setup_notation}

Suppose longitudinal data are observed from \(n\) independent subjects. For subject \(i\), \(i=1,\ldots,n\), with \(n_i\) repeated observations, let \(Y_{ij}\) denote the response measured at time \(t_{ij}\), where \(j=1,\ldots,n_i\) indexes the observed time points for that subject. The full dataset is denoted by
\[
\mathcal{D}=\{(i,t_{ij},Y_{ij}): i=1,\ldots,n,\ j=1,\ldots,n_i\}.
\]

The goal is to detect time points at which the slope of the temporal trajectory of \(Y_{ij}\) changes. For \(K\) change points \(\tau_1,\ldots,\tau_K\), a general slope-change working model can be written as
\begin{equation}
g\{E(Y_{ij}\mid t_{ij})\}
=
\beta_0+\beta_1 t_{ij}
+
\sum_{k=1}^{K}\beta_k(t_{ij}-\tau_k)_+,
\label{eq:slope_change_model}
\end{equation}
where \((t_{ij}-\tau_k)_+=\max(t_{ij}-\tau_k,0)\), \(g(\cdot)\) is a link function or the identity link depending on the outcome type, and \(\beta_k\) represents the slope change at \(\tau_k\). For quantile-regression-based implementations, the same linear predictor is used to model the conditional quantile of \(Y_{ij}\).

This formulation is flexible and can be implemented using different working models, as described in Section~\ref{sec:working_models}. For longitudinal data with within-subject dependence, generalized least squares or mixed-effects models can be used. For data with outliers or non-normal errors, quantile regression provides a robust alternative, and quantile rank-score statistics further allow within-subject dependence to be incorporated through a correlation-adjusted variance estimator, as described in Section~\ref{sec:quantile_rank_score}.

For a given time segment, observation times are rescaled to the interval \([0,1]\), where 0 and 1 denote the beginning and end of the segment, respectively. Let
\[
\mathcal{T}=\{c_1,\ldots,c_M\}
\]
denote the candidate change-point set within the segment, with \(0<c_1<\cdots<c_M<1\). For each candidate \(c_m\), let \(T_m\) denote the corresponding test statistic measuring evidence for a slope change. The largest candidate-specific statistic is taken as the observed statistic for the segment, and its corresponding candidate location is selected as the estimated change-point location:
\[
T_{\mathrm{obs}}=\max_{1\leq m\leq M}T_m,
\qquad
\widehat{c}=c_{\arg\max_{1\leq m\leq M}T_m}.
\]
The statistical significance of \(T_{\mathrm{obs}}\) is then evaluated using the Loop Permutation Test. If the resulting permutation \(p\)-value is smaller than the prespecified significance level \(\alpha\), \(\widehat{c}\) is retained as a detected change point in the current segment. Otherwise, no significant change point is detected in that segment. The details of this procedure are described in Section~\ref{sec:loop_permutation}.

For multiple change-point detection, we apply the single-change-point Loop Permutation Test (Section~\ref{sec:loop_permutation}) recursively using binary segmentation, as described in Section~\ref{sec:binary_segmentation}. Let
\[
\widehat{C}=\{\widehat{c}_1,\ldots,\widehat{c}_{\widehat{K}}\}
\]
denote the estimated set of change points obtained from this procedure, where \(\widehat{K}\) is the estimated number of change points. The significance level used to determine whether a candidate change point is retained is denoted by \(\alpha\), and the number of permutations used to construct the empirical null distribution is denoted by \(B\).

\subsection{Loop Permutation Test}
\label{sec:loop_permutation}

The Loop Permutation Test is motivated by cyclic-shift and block-resampling ideas that preserve local dependence while disrupting global alignment. In genomic studies, cyclic-shift permutation has been used to test recurrent genomic aberrations by circularly shifting ordered genomic measurements within each sample, thereby preserving local correlation among nearby markers while breaking cross-sample alignment of aberrant regions \citep{walter2015consistent}. Related circular genomic permutation ideas have also been used in GWAS gene-set and network analyses to preserve genomic structure while generating an empirical null distribution \citep{cabrera2012uncovering}. More generally, resampling blocks or circular segments to preserve temporal dependence is well established in the bootstrap literature \citep{kunsch1989jackknife,politis1994stationary,davison1997bootstrap}. Following this principle, the proposed Loop Permutation Test applies circular shifts to null-model residual trajectories within each subject. Circularly shifting these residuals preserves local within-subject dependence while breaking their alignment with the original time points, producing a finite-sample permutation null distribution for testing slope changes.

Within a given segment, the Loop Permutation Test first scans all candidate change points in \(\mathcal{T}\). For each candidate \(c_m\), \(m=1,\ldots,M\), the slope-change model in Equation~\eqref{eq:slope_change_model} is fitted with \(K=1\) and \(\tau_1=c_m\). Evidence for a change point at \(c_m\) is summarized by \(T_m\), which may be based on \(-\log_{10}(p\text{-value})\), a score-test statistic, a \(z\)- or \(t\)-statistic, or a log-likelihood-ratio statistic. The candidate with the largest statistic is selected as \(\widehat{c}\), and \(T_{\mathrm{obs}}\) is used as the observed evidence for a slope change in the segment.

To assess significance, a null model without a change point is first fitted, and the residual trajectory for each subject is obtained under this null model. For each permutation \(b=1,\ldots,B\), the residual trajectory is circularly shifted within subject to obtain \(r_{ij}^{(b)}\). For example, if subject \(i\) has residual trajectory
\[
(r_{i1}, r_{i2}, \ldots, r_{in_i}),
\]
then a circular shift by two observed time points gives
\[
(r_{i3}, r_{i4}, \ldots, r_{in_i}, r_{i1}, r_{i2}).
\]
This operation is performed independently for each subject, preserving the within-subject residual ordering while disrupting its alignment with the original time points. A permuted response is then generated by adding the shifted residuals back to the null-model fitted values:
\[
Y_{ij}^{(b)}=\widehat{Y}_{ij}+r_{ij}^{(b)}.
\]
For each permuted dataset, the candidate scan over \(\mathcal{T}\) is repeated, and the maximum statistic is recorded as \(T^{(b)}\). For the quantile rank-score implementation, the same circular-shift idea is applied to the null quantile residual trajectory, and the rank-score statistic is recomputed using the permuted residual signs. The permutation \(p\)-value is computed as
\[
p =
\frac{
1+\sum_{b=1}^{B} I\{T^{(b)} \geq T_{\mathrm{obs}}\}
}{
B+1
}.
\]
The Loop Permutation Test returns the estimated change point \(\widehat{c}\), the observed statistic \(T_{\mathrm{obs}}\), and the permutation \(p\)-value. 

\begin{figure}[!htbp]
    \centering
    \includegraphics[width=\textwidth]{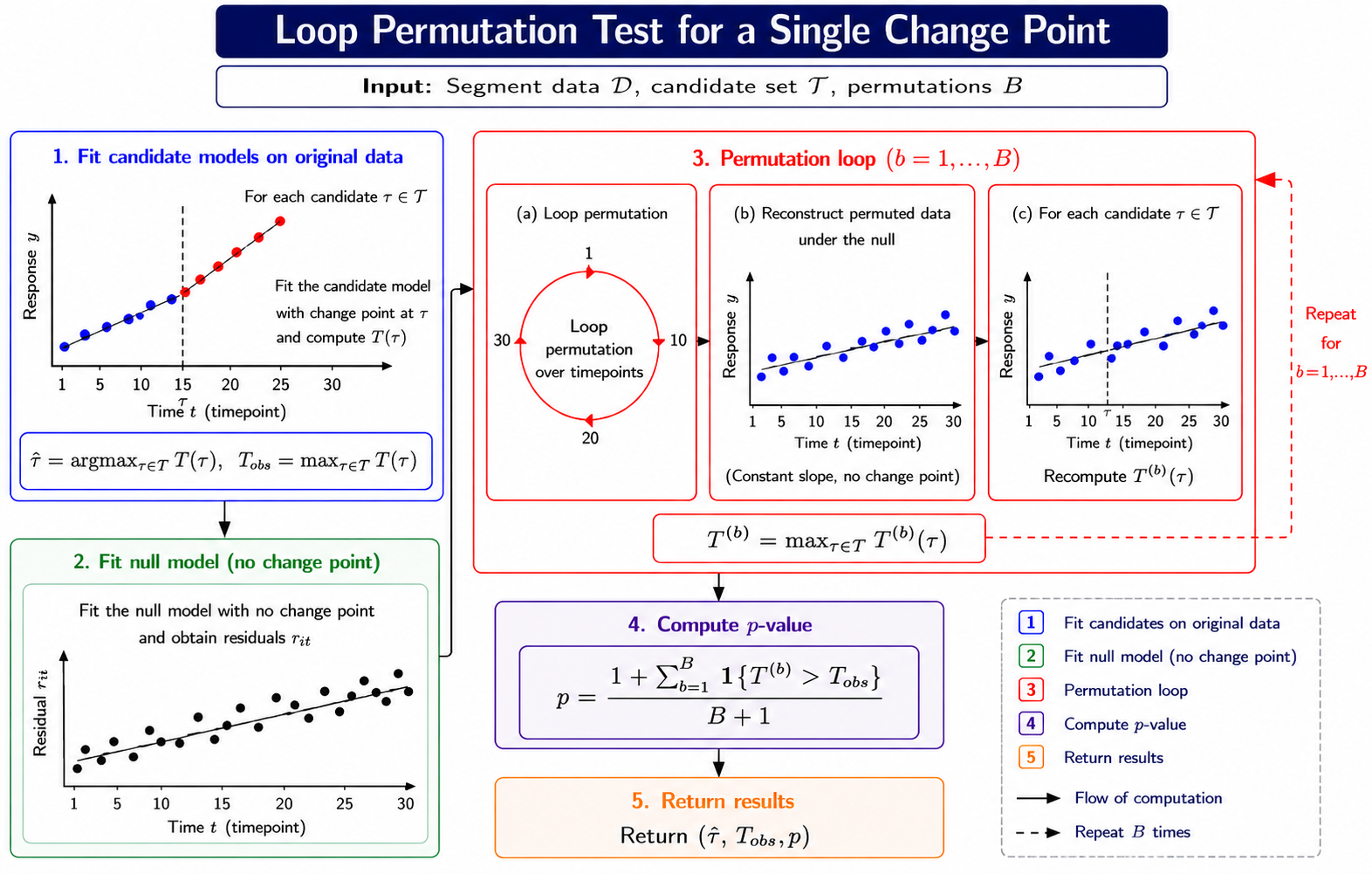}
    \caption{Overview of the Loop Permutation procedure for change-point detection in longitudinal data.}
    \label{fig:loop_permutation}
\end{figure}

\subsection{Binary Segmentation for Multiple Change Points}
\label{sec:binary_segmentation}
\begin{figure}[!htbp]
    \centering
    \includegraphics[width=\textwidth]{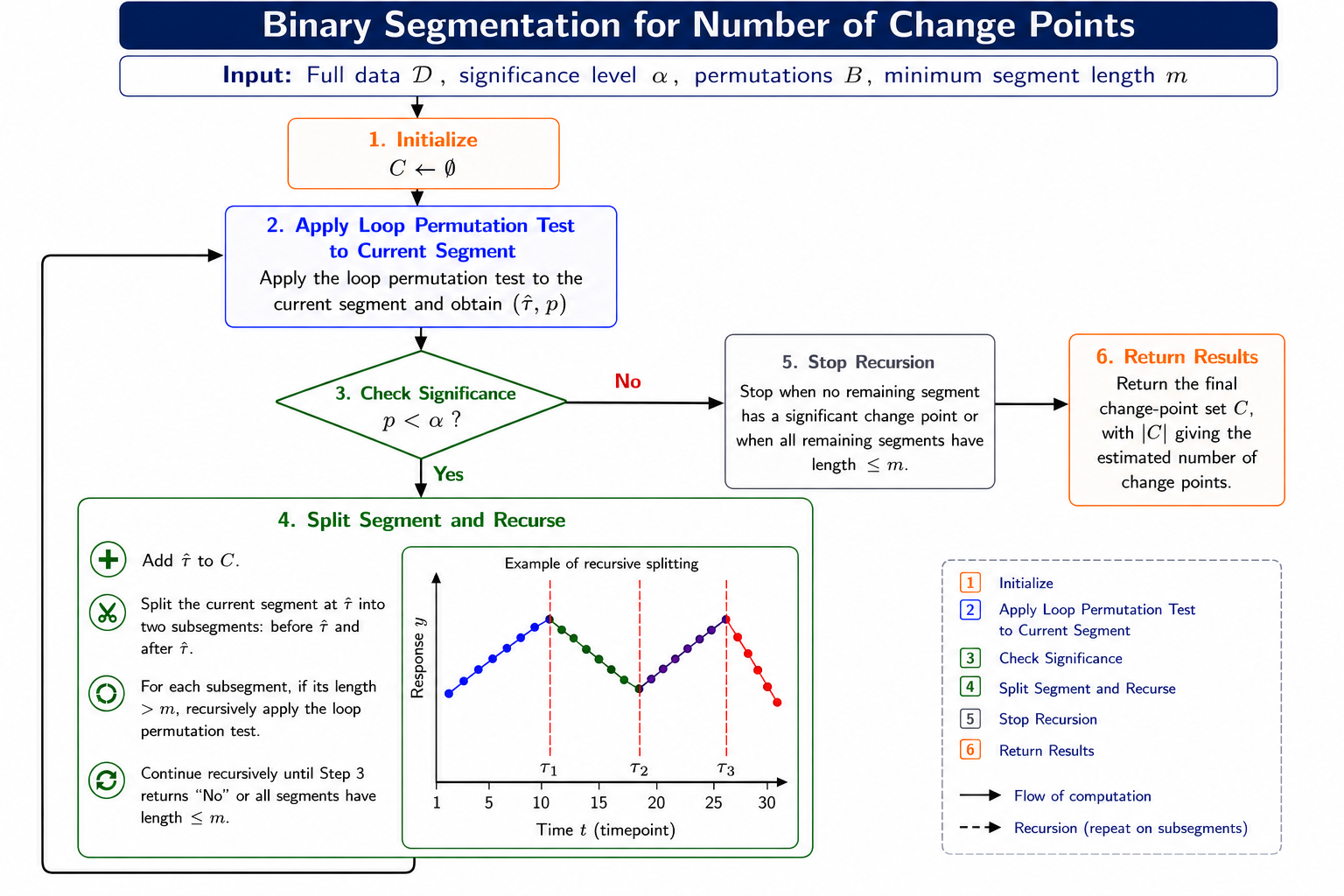}
    \caption{Overview of the binary segmentation procedure for detecting multiple change points.}
    \label{fig:binary_segmentation}
\end{figure}

To estimate the number and locations of multiple change points (Figure~\ref{fig:binary_segmentation}), binary segmentation is used. If a change point is detected in a segment, the loop permutation procedure described above is repeated separately on the left and right subsegments defined by the detected change point, recursively searching for additional change points in each.
The search stops when no new change points are detected in any remaining subsegment. 







\subsection{Working Models}
\label{sec:working_models}

The Loop Permutation framework can be implemented with flexible working models and test statistics. For longitudinal data, models that account for within-subject dependence can be used, including mixed-effects models, generalized estimating equations, and generalized least squares models. These mean-based approaches are standard, so their estimating equations and test statistics are not repeated here. In the motivating longitudinal genomic applications, measurements are often sparse and irregularly timed, and expression profiles may contain outliers or depart from normality. Under these conditions, mean-based inference can be sensitive to extreme observations. To provide a more robust alternative, a quantile-based Loop Permutation Test is also implemented. In particular, to incorporate within-subject dependence within the quantile framework, the quantile rank-score test with a correlation-adjusted variance estimator is used \citep{wang2008enhanced_quantile}. Because this component is less standard and is central to the robust version of the proposed method, the quantile rank-score formulation is described in detail in Section~\ref{sec:quantile_rank_score}.

\subsubsection{Quantile Rank-Score Test}
\label{sec:quantile_rank_score}

The quantile rank-score test proposed by \citet{wang2008enhanced_quantile} provides a robust test statistic for longitudinal quantile regression while accounting for within-subject dependence. For a given quantile level \(q\), a null quantile regression model without a change point is first fitted to obtain residuals \(r_{ij}\). For each candidate change point \(c_m\in\mathcal{T}\), the corresponding slope-change covariate is orthogonalized with respect to the null-model covariates, and the resulting covariate is denoted by \(z^*_{ij}(c_m)\). The quantile rank-score statistic is
\[
S_n(c_m)=n^{-1/2}\sum_{i,j} z^*_{ij}(c_m)\psi_q(r_{ij}),
\qquad
\psi_q(u)=q-I(u<0).
\]

To account for within-subject dependence, the variance is estimated using a sign-correlation adjustment:
\[
Q_n(\widehat{\delta};c_m)
=
\frac{1}{n}\sum_{i,j}\{z^*_{ij}(c_m)\}^2q(1-q)
+
\frac{1}{n}\sum_i\sum_{j_1\neq j_2}
z^*_{ij_1}(c_m)z^*_{ij_2}(c_m)(-q^2+\widehat{\delta}),
\]
where
\[
\widehat{\delta}
=
(L-p)^{-1}
\sum_i \sum_{j_1\neq j_2}
I\{r_{ij_1}<0,\ r_{ij_2}<0\}.
\]
Here, \(L\) is the total number of within-subject observation pairs, and \(p\) is the number of nuisance parameters in the null quantile regression model. The second term in \(Q_n(\widehat{\delta};c_m)\) adjusts for dependence among repeated observations from the same subject.

The resulting candidate-specific statistic is
\[
T_m=\frac{S_n^2(c_m)}{Q_n(\widehat{\delta};c_m)}.
\]
This \(T_m\) corresponds to the candidate-specific statistic defined in Section~\ref{sec:setup_notation}. 
\section{SIMULATION STUDY}

The proposed method was evaluated under several simulation scenarios. First, Type I error control was assessed under the null setting with no true change point. Second, detection performance was evaluated under settings with one or two true change points. The effects of key factors were examined, including sample size, number of time points, outlier proportion, missing-data rate, and effect size. Outlier scenarios were included to mimic real longitudinal data with extreme observations, while missing-data scenarios were used to mimic irregular observation patterns in which subjects may have different observed time points.

\subsection{Simulation Settings}

Longitudinal data were generated for \(n\) independent subjects observed at \(J\) time points. For subject \(i\) at time \(t\), the response was simulated as

\[
Y_{it}=\mu(t)+\varepsilon_{it},
\]

where the within-subject errors followed an AR(1) process,

\[
\varepsilon_{it}=\rho\varepsilon_{i,t-1}+\eta_{it}, 
\qquad 
\eta_{it}\sim N\{0,(1-\rho^2)\sigma^2\},
\]

with \(\varepsilon_{i1}\sim N(0,\sigma^2)\). The mean function \(\mu(t)\) was piecewise linear:
\[
\mu(t)=\alpha_0+\alpha_1 t+\sum_{k=1}^{K}\beta_k(t-\tau_k)_+,
\]
where \(K=0,1,\) or \(2\) denotes the number of true change points, \((t-\tau_k)_+=\max(t-\tau_k,0)\), and \(\beta_k\) represents the slope change at \(\tau_k\). When \(K=0\), this reduces to the no-change-point model \(\mu(t)=\alpha_0+\alpha_1 t\). Change points were sampled from an interior time range, with a minimum spacing requirement when multiple change points were present. Outliers were generated by randomly selecting a proportion of observations and adding large mean-zero Gaussian noise. The proposed GLS-, quantile-regression-, and rank-score-based Loop Permutation methods were compared with \texttt{ecp} \citep{james2015ecp}, \texttt{changepoint} \citep{killick2014changepoint}, \texttt{EnvCpt} \citep{killick2021envcpt}, and \texttt{strucchange} \citep{zeileis2002strucchange}, using \(B=500\) permutations and significance level \(\alpha=0.05\).

Simulation scenarios varied sample size, number of time points, outlier proportion, and missing-data rate. Specifically, sample sizes included \(n=25,50,100,\) and \(200\); the numbers of time points included \(J=20,30,40,\) and \(50\); outlier proportions included \(0,0.10,0.25,\) and \(0.50\); and missing-data rates included \(0\%,10\%,20\%,\) and \(40\%\). Effect size was defined as the slope change at the true change point. For the one-change-point setting, effect sizes of \(0.1, 0.2, 0.3,\) and \(0.5\) were considered. For the two-change-point setting, both change points were assigned the same effect size, with values \(0.25, 0.5, 1.5,\) and \(2\). In each scenario, one factor was varied while the remaining factors were held fixed at their default values.

\subsection{Type I Error Control}

Type I error was assessed under the null setting with no true change point. For each simulated dataset, a false positive was defined as detecting any change point. The Type I error rate was then calculated as the proportion of null datasets in which at least one change point was detected. The proposed LoopPerm\_gls, LoopPerm\_qr, and LoopPerm\_rank\_score methods controlled Type I error near the nominal 0.05 level across different sample sizes, outlier proportions, numbers of time points, and missing-data rates. In contrast, \texttt{ecp} and \texttt{changepoint} were overly conservative, whereas \texttt{strucchange} showed inflated Type I error and \texttt{EnvCpt} showed inflation when the number of time points was small.

\begin{figure}[!htbp]
    \centering
    \includegraphics[width=\textwidth]{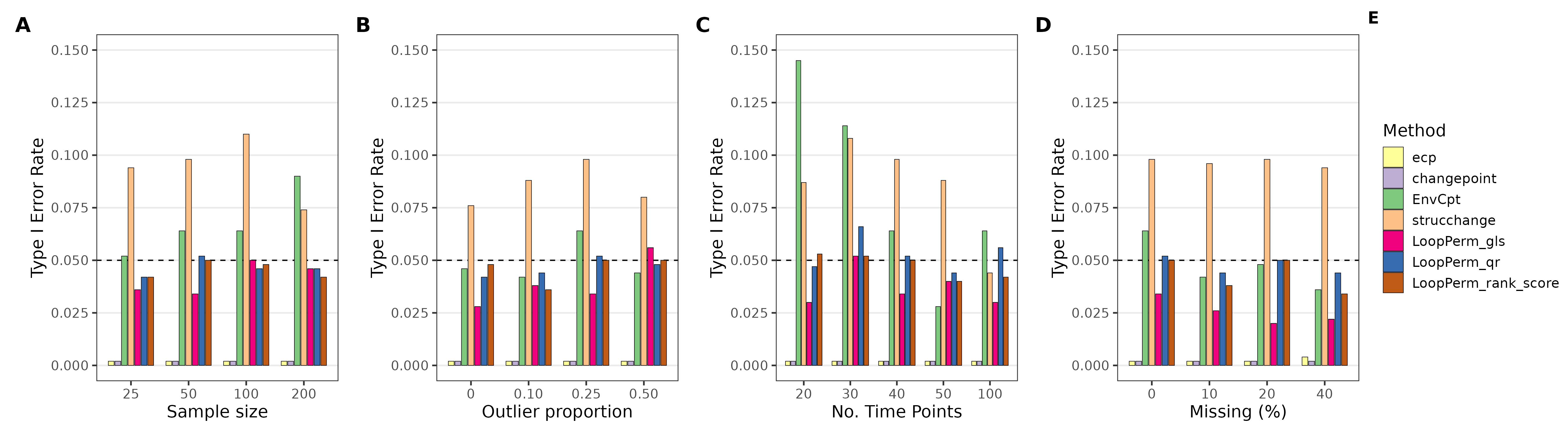}
    \caption{Type I error rates across methods under the null setting with no true change point. The default setting used \(n=40\), \(J=40\) time points, an outlier proportion of 0.25, and missing rate 0; each panel varied one factor while holding the others fixed at these default values.}
    \label{fig:type1_panels}
\end{figure}
\subsection{Power}

Power was evaluated under one-change-point and two-change-point settings. A detection was considered correct only when the method identified the exact number of true change points and each estimated change point was within 3 time units of the corresponding true location. Overall, the proposed LoopPerm methods achieved higher power than the competing methods across most scenarios. In the one-change-point setting, LoopPerm\_gls generally performed best, with power increasing as sample size, number of time points, and effect size increased. In the two-change-point setting, LoopPerm\_qr and LoopPerm\_rank\_score showed better performance than LoopPerm\_gls, indicating improved robustness for more complex change-point structures. Under scenarios with outliers, the nonparametric LoopPerm\_qr and LoopPerm\_rank\_score methods were more robust than the GLS-based method. Most competing methods showed lower or less stable power across settings.

\begin{figure}[!htbp]
    \centering

    \includegraphics[width=\textwidth]{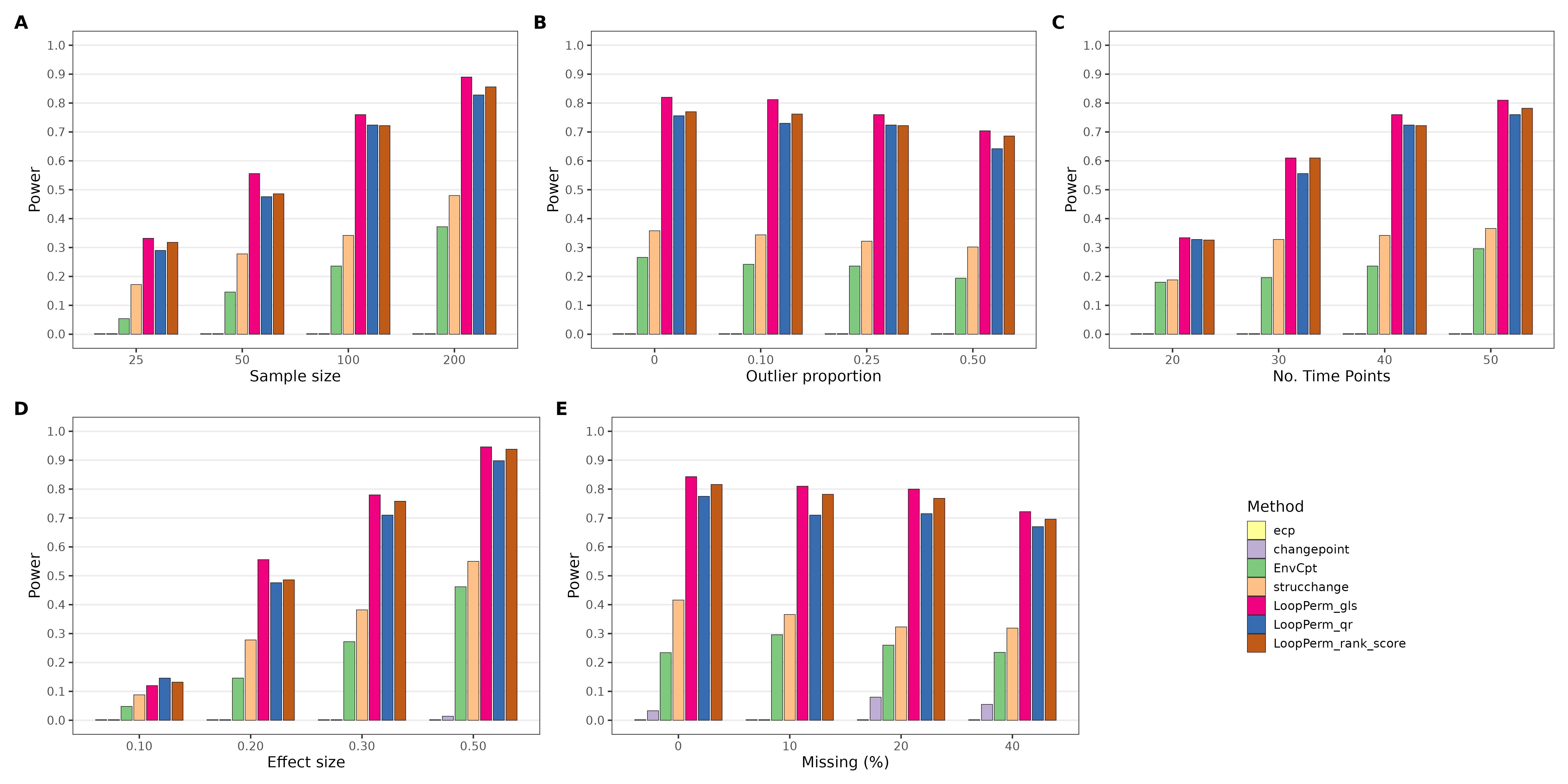}

    \vspace{0.4cm}

    \includegraphics[width=\textwidth]{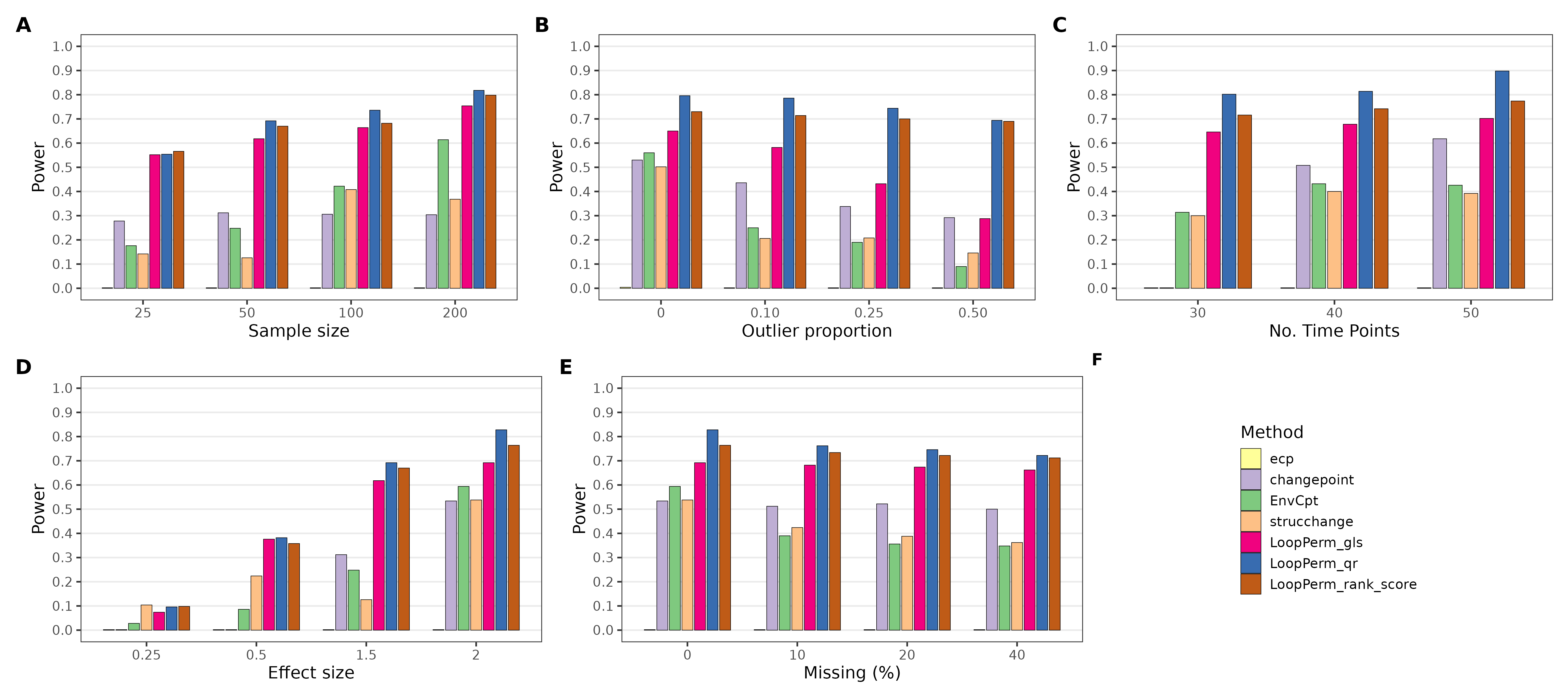}

    \caption{Power comparison across methods under the one-change-point and two-change-point settings. Top: one-change-point setting; bottom: two-change-point setting. For the one-change-point setting, all scenarios except the sample-size analysis used \(n=50\), with default outlier proportion 0.25, \(J=40\) time points, effect size 0.2, and missing rate 0. For the two-change-point setting, the two change points had the same effect size, and the default setting used \(n=50\), \(J=40\), outlier proportion 0.25, effect size 1.5, and missing rate 0. In each panel, one factor was varied while the others were held fixed at their default values.}
    \label{fig:power_comparison}
\end{figure}

\section{DATA APPLICATION}
The proposed method was further applied to longitudinal transcriptomic data from the H1N1 and H3N2 human viral inoculation studies in GSE73072 \citep{liu2016individualized}. The analysis focused on two types of longitudinal immune features: xCell-derived cell-type enrichment scores \citep{aran2017xcell} and selected single-gene expression trajectories. xCell scores are transcriptome-based estimates of the relative enrichment of immune cell types inferred from bulk gene-expression data. Because both the H1N1 and H3N2 datasets included samples from two centers, batch-effect correction was applied before conducting change-point analysis to account for center-related technical variation. Gene-expression data were first quantile normalized \citep{bolstad2003comparison}, after which \texttt{limma} linear models were fitted using baseline pre-inoculation samples with study center as a covariate \citep{ritchie2015limma}. The resulting batch-effect estimates were then applied to the full longitudinal dataset, including post-inoculation time points, to generate corrected expression profiles while retaining biologically meaningful temporal patterns.

\begin{figure}[!htbp]
    \centering

    \begin{subfigure}{0.95\textwidth}
        \centering
        \includegraphics[width=\textwidth]{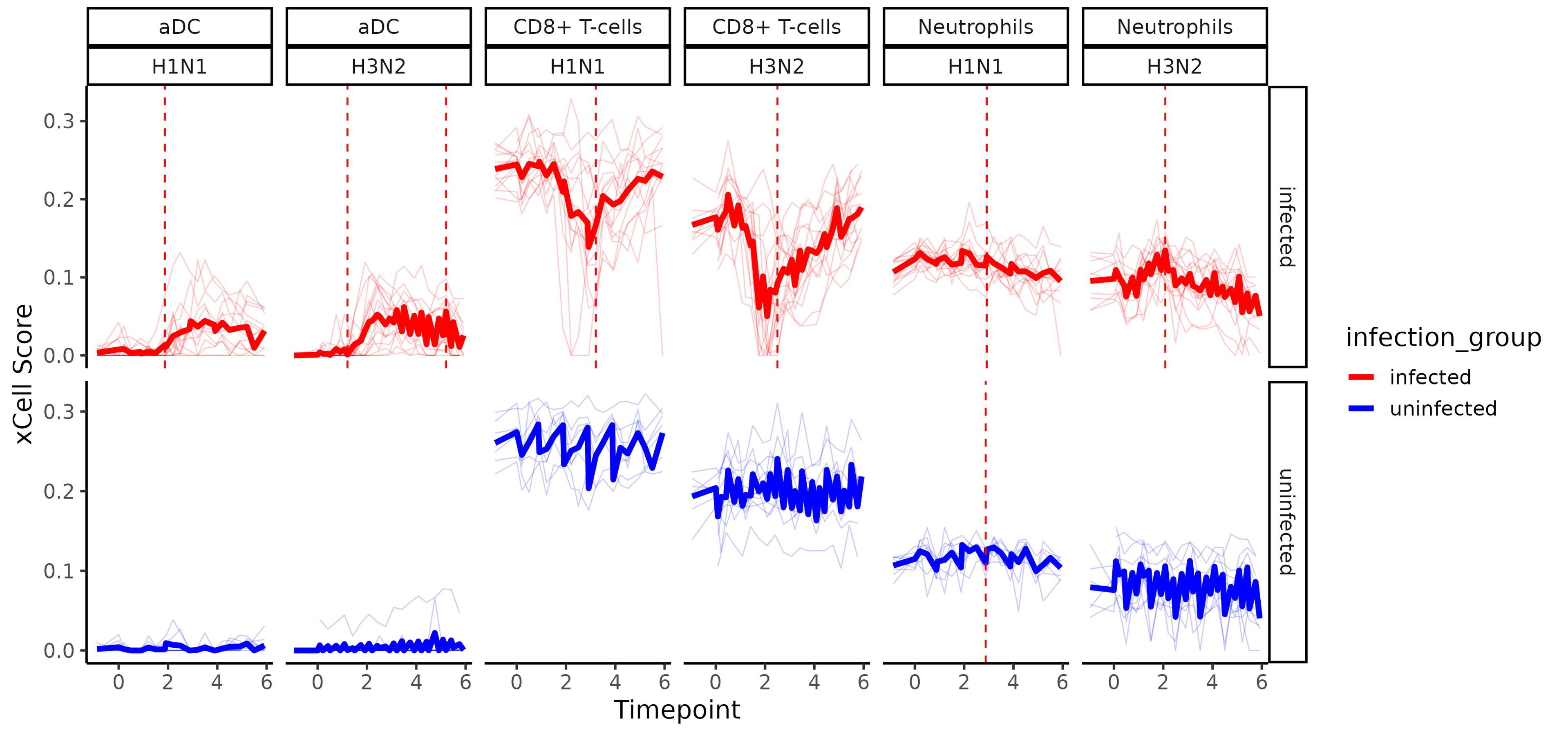}
        \caption{xCell score trajectories over time.}
        \label{fig:xcell_panel}
    \end{subfigure}

    \vspace{0.5cm}

    \begin{subfigure}{0.95\textwidth}
        \centering
        \includegraphics[width=\textwidth]{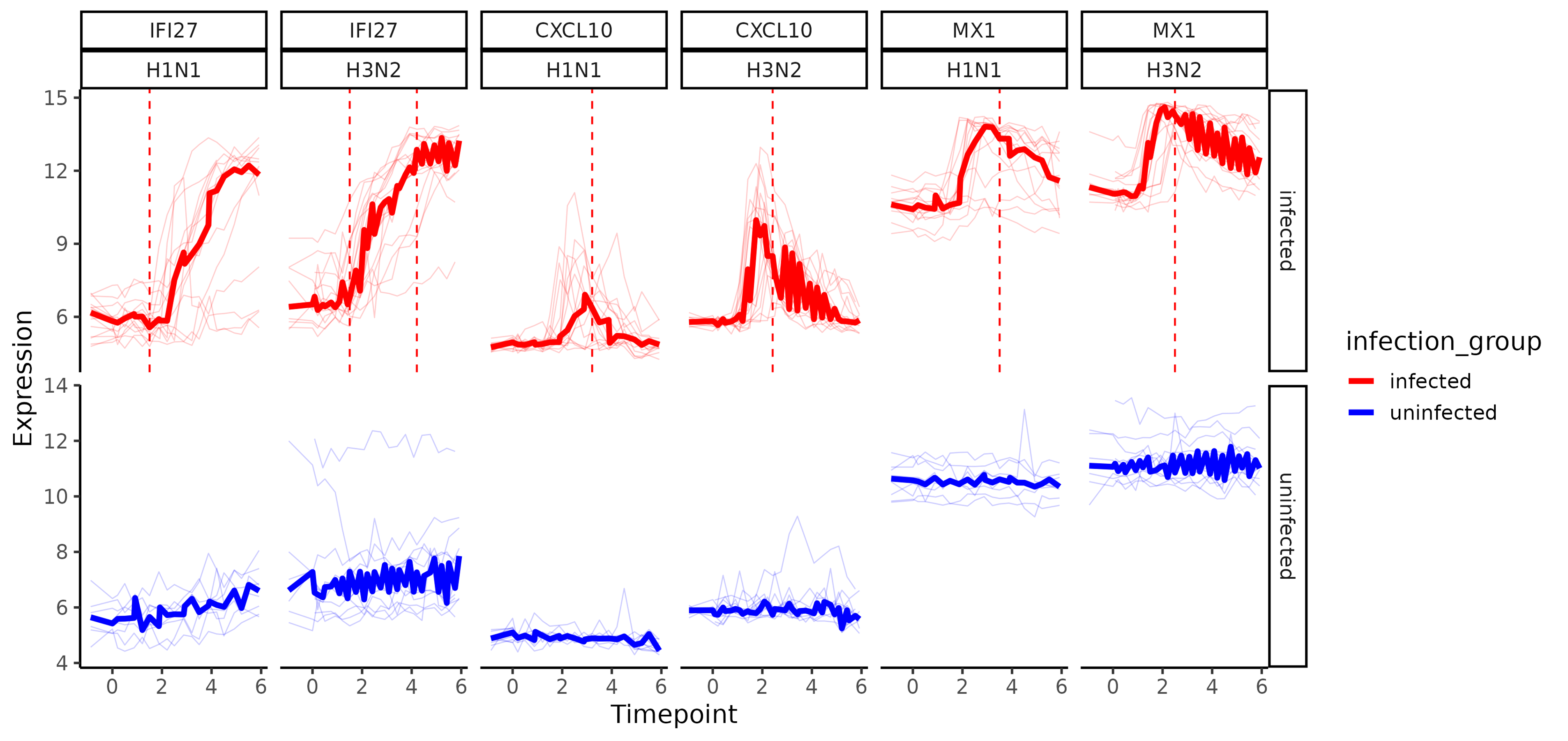}
        \caption{Gene-expression trajectories over time.}
        \label{fig:gene_expression_panel}
    \end{subfigure}

    \caption{Longitudinal cell-type score and gene-expression applications of the proposed change-point detection method.}
    \label{fig:xcell_gene_applications}
\end{figure}

Subjects were classified into two groups, infected/symptomatic and uninfected/asymptomatic, based on laboratory records and symptoms after inoculation. Change-point detection was then applied separately to each group. Figure~\ref{fig:xcell_gene_applications} shows the detected change points for selected xCell scores and single-gene expression trajectories. Overall, infected/symptomatic subjects showed more pronounced temporal changes, whereas uninfected/asymptomatic subjects remained relatively stable over time. In the xCell enrichment analysis, early changes were primarily observed in innate immune-related cell types, including activated dendritic cells (aDCs) and neutrophils, whereas later changes in CD8\textsuperscript{+} T cells suggested subsequent adaptive immune activation. In the single-gene trajectory analysis, IFI27, MX1, and CXCL10 showed pronounced temporal activation patterns consistent with interferon-mediated antiviral responses, with IFI27 exhibiting particularly early transcriptional changes. Compared with H1N1, H3N2 generally exhibited earlier change points across both cell-type and gene-expression trajectories, suggesting a more rapid host transcriptional response following inoculation. These findings are consistent with experimental studies showing early innate immune responses and rapid interferon activation after influenza viral challenge.

\section{CONCLUSION AND DISCUSSION}
\label{s:discuss}
This study proposes Loop Permutation as a flexible framework for automatic multiple change-point detection in longitudinal data. The key methodological innovation is the within-subject circular permutation of null-model residual trajectories, which generates a permutation-based null distribution while preserving the local dependence structure of repeated measurements. By combining the Loop Permutation Test with binary segmentation, the proposed framework estimates both the number and locations of change points without requiring them to be specified in advance. The framework can be implemented with flexible modeling choices, including GLS-based models, quantile regression, and rank-score-based test statistics. Simulation studies showed that Loop Permutation controlled Type I error near the nominal level and generally achieved higher power than competing methods, with the nonparametric LoopPerm\_qr and LoopPerm\_rank\_score methods showing stronger performance and robustness in more complex or outlier-contaminated settings. Real-data applications to H1N1/H3N2 inoculation transcriptomic data further demonstrated that Loop Permutation can identify interpretable transition points in longitudinal genomic trajectories.

Loop Permutation has broad potential applications in longitudinal biomedical studies. In longitudinal genomics, it can be used to identify key transition times in vaccination, infection, treatment-response, and developmental studies. More broadly, Loop Permutation can be applied to repeated-measures clinical biomarkers, imaging-derived features, wearable-device measurements, and population health trajectories when the timing of trajectory changes is of interest. Because Loop Permutation provides permutation-based inference for detected change points, it can help prioritize statistically supported transition times for biological or clinical interpretation.

Several limitations remain. First, because Loop Permutation is permutation-based, it can be computationally intensive, especially for long time series, large candidate grids, or a large number of permutations. Second, the method requires prespecified choices for the candidate change-point grid and minimum segment length, which may influence performance in settings with sparse or irregular observation times. Third, the current implementation focuses mainly on univariate slope changes, whereas many omics studies involve high-dimensional features and coordinated multivariate changes across genes, pathways, or cell-type scores. Finally, binary segmentation detects change points sequentially rather than simultaneously; therefore, the first detected change point is not necessarily the earliest in time, but rather the one with the strongest evidence for a slope change within the current segment.

Future work will extend Loop Permutation in several directions. Parallel or simultaneous detection strategies could be developed to identify multiple change points more efficiently and reduce reliance on sequential binary segmentation. Computational acceleration, including parallelized permutations and more efficient candidate scanning, would improve scalability for longer time series and larger datasets. Extensions to multivariate or high-dimensional longitudinal data would further allow Loop Permutation to detect coordinated transition points across genes, pathways, or cell-type scores.


\backmatter


\section*{Funding}
This work was supported in part by the National Institutes of Health under grant R01 AI170959-01A1.

\section*{Data availability}

The data underlying this article are publicly available from the Gene Expression Omnibus (GEO). The influenza challenge data used in the real-data application can be accessed through GEO under accession number GSE73072 (\url{https://www.ncbi.nlm.nih.gov/geo/query/acc.cgi?acc=GSE73072}).

\providecommand{\newblock}{}
\bibliographystyle{biom} 
\bibliography{mybibilo}


\clearpage
\appendix

\section{Algorithm Details}

The algorithms below provide formal pseudocode for the procedures illustrated in Figures~\ref{fig:loop_permutation} and~\ref{fig:binary_segmentation}. Algorithm~\ref{alg:loop_permutation} corresponds to the Loop Permutation Test for a single change point within a given segment, and Algorithm~\ref{alg:binary_segmentation} corresponds to the binary segmentation procedure for detecting multiple change points. The figures provide a visual overview of the workflow, whereas the algorithms present the same procedures in step-by-step computational form.

\begin{algorithm}[t]
\caption{Loop Permutation Test for a Single Change Point}
\label{alg:loop_permutation}
\begin{algorithmic}
\State \textbf{Input:} Segment data \(\mathcal{D}_s\), candidate set \(\mathcal{T}_s=\{c_1,\ldots,c_M\}\), number of permutations \(B\), minimum segment length \(m\)
\State \textbf{Output:} Estimated change point \(\widehat{c}\), observed statistic \(T_{\mathrm{obs}}\), permutation \(p\)-value \(p\)

\If{segment length of \(\mathcal{D}_s \leq m\)}
    \State \Return \((\mathrm{NA}, \mathrm{NA}, 1)\)
\EndIf

\For{\(c_m \in \mathcal{T}_s\), \(m=1,\ldots,M\)}
    \State Fit the candidate slope-change model with \(K=1\) and \(\tau_1=c_m\)
    \State Compute the candidate-specific statistic \(T_m\)
\EndFor

\State \(T_{\mathrm{obs}} \gets \max_{1\leq m\leq M} T_m\)
\State \(\widehat{c} \gets c_{\arg\max_{1\leq m\leq M}T_m}\)

\State Fit the null model without a change point
\State Obtain fitted values \(\widehat{Y}_{ij}\) and residuals \(r_{ij}=Y_{ij}-\widehat{Y}_{ij}\)

\For{\(b = 1,\ldots,B\)}
    \State Circularly shift residuals within each subject to obtain \(r_{ij}^{(b)}\)
    \State Generate the permuted response \(Y_{ij}^{(b)}=\widehat{Y}_{ij}+r_{ij}^{(b)}\)

    \For{\(c_m \in \mathcal{T}_s\), \(m=1,\ldots,M\)}
        \State Fit the candidate slope-change model to the permuted data
        \State Compute the permuted candidate statistic \(T_m^{(b)}\)
    \EndFor

    \State \(T^{(b)} \gets \max_{1\leq m\leq M} T_m^{(b)}\)
\EndFor

\State Compute the permutation \(p\)-value:
\[
p =
\frac{
1+\sum_{b=1}^{B}\mathbf{1}\{T^{(b)} \geq T_{\mathrm{obs}}\}
}{
B+1
}.
\]

\State \Return \((\widehat{c}, T_{\mathrm{obs}}, p)\)
\end{algorithmic}
\end{algorithm}

\begin{algorithm}[t]
\caption{Binary Segmentation with Loop Permutation}
\label{alg:binary_segmentation}
\begin{algorithmic}
\State \textbf{Input:} Full data \(\mathcal{D}\), significance level \(\alpha\), number of permutations \(B\), minimum segment length \(m\)
\State \textbf{Output:} Detected change-point set \(\widehat{C}\)

\State Initialize \(\widehat{C} \gets \emptyset\)

\Function{BinarySegmentation}{$\mathcal{D}_s$}
    \If{segment length of \(\mathcal{D}_s \leq m\)}
        \State Stop splitting this segment
    \EndIf

    \State Rescale observation times in \(\mathcal{D}_s\) to \([0,1]\)
    \State Construct candidate set \(\mathcal{T}_s=\{c_1,\ldots,c_M\}\) by excluding boundary time points
    \State Apply Algorithm~\ref{alg:loop_permutation} to \(\mathcal{D}_s\) with candidate set \(\mathcal{T}_s\)
    \State Obtain \((\widehat{c}, T_{\mathrm{obs}}, p)\)

    \If{\(p \leq \alpha\)}
        \State Add \(\widehat{c}\) to \(\widehat{C}\)
        \State Split \(\mathcal{D}_s\) into left and right subsegments at \(\widehat{c}\)
        \State Recursively apply \Call{BinarySegmentation}{left subsegment}
        \State Recursively apply \Call{BinarySegmentation}{right subsegment}
    \Else
        \State Stop splitting this segment
    \EndIf
\EndFunction

\State Apply \Call{BinarySegmentation}{$\mathcal{D}$}
\State \Return sorted \(\widehat{C}\)

\end{algorithmic}
\end{algorithm}

\label{lastpage}
\end{document}